\documentclass[useAMS,usenatbib]{mn2e}
\usepackage{graphicx,fleqn}
\DeclareGraphicsExtensions{.eps,.ps,.eps.gz,.ps.gz,.eps.Z}
\def \zeroseven{RX J0720.4-3125}
\def \zerofour{RX J0420.0-5022}

\def \apj{ApJ}
\def \apjl{ApJL}

\def \aap{A\&A}
\def \aaps{A\&A Suppl.}
\def \mnras{MNRAS}

\def \pasp{PASP}

\input psfig.sty

\title[Unveiling the thermal and magnetic map of neutron star 
surfaces...]{
Unveiling the thermal and magnetic map of neutron star surfaces though
their X-ray emission: method and lightcurves analysis}

\author[]{Silvia Zane$^{1}$ and Roberto Turolla$^{2}$ \\
$^{1}$Mullard Space Science Laboratory, University College London, Holmbury St. Mary,
Dorking Surrey, RH5 6NT, UK; sz@mssl.ucl.ac.uk \\
$^{2}$Department of Phisics, University of Padua, via Marzolo 8, I-35131, Padova, Italy;
turolla@pd.infn.it
}

\begin{document}

\date{Accepted...Received...}


\maketitle

\label{firstpage}

\begin{abstract}

Recent {\em Chandra} and {\em XMM-Newton} observations of a number of
X-ray ``dim'' pulsating neutron stars revealed quite unexpected features
in the emission from these sources. Their soft thermal spectrum, believed
to originate directly from the star surface, shows evidence for a
phase-varying absorption line at some hundred eVs. The pulse modulation is
relatively large (pulsed fractions in the range $\sim 12\%$--$35\%$), the
pulse shape is often non-sinusoidal, and the hard X-ray color appears to
be anti-correlated in phase with the total emission. Moreover, the
prototype of this class, \zeroseven, has been found to undergo rather
sensible changes both in its spectral and timing properties over a
timescale of a few years. All these new findings seem difficult to
reconcile with the standard picture of a cooling neutron star endowed with
a purely dipolar magnetic field, at least if surface emission is produced
in an atmosphere on top of the crust. In this paper we explore how a
dipolar+quadrupolar star-centered field influence the properties of the
observed lightcurves. The phase-resolved spectrum has been evaluated
accounting for both radiative transfer in a magnetized atmosphere and
general relativistic ray-bending. We computed over 78000 lightcurves
varying the quadrupolar components and the viewing geometry. A comparison
of the data with our model indicate that higher order multipoles are
required to reproduce the observations.

\end{abstract}


\section{INTRODUCTION}
\label{intro}

Over the last few years a number of high resolution spectral and timing
observations of thermally emitting neutron stars (NSs) have become available
thanks to new generation X-ray satellites (both {\it Chandra\/} and {\it XMM-Newton\/}),
opening new perspectives in the study of these sources. Thermal emission from
isolated NSs is presently observed in more than 20 sources, including
active radio pulsars, soft $\gamma$-repeaters, anomalous X-ray pulsars,
Geminga and Geminga-like objects, and X-ray dim radio-silent NSs. There is
by now a wide consensus that the soft, thermal component directly
originates from the surface layers as the star cools down. If properly exploited,
the information it conveys are bound to reveal much about the physics of neutron stars,
shedding light on their thermal and magnetic surface distribution and ultimately
probing the equation of state of matter at supra-nuclear densities.

Although thermal surface emission seems indeed to be an ubiquitous feature in isolated
NSs, a power-law, non-thermal component (likely produced in the star magnetosphere) is
present in most sources, where it often dominates the X-ray spectrum. Moreover, the
intrinsic X-ray emission from young radio-pulsars may be significantly contaminated
by the contribution of the surrounding supernova remnant. In this respect
the seven dim X-ray sources discovered by {\it ROSAT\/} (hereafter XDINSs) are a most
notable exception. In a sense, one may claim that these are the only ``genuinely isolated''
NSs and their soft thermal emission is unmarred by (non-thermal) magnetospheric
activity nor by the presence of a supernova remnant or a binary companion (see
e.g. \citealt{t2000} and \citealt{hab2004} for reviews; \citealt{sil05}). XDINSs play a
key role in compact objects astrophysics: these are the only sources in which we can have a
clean view of the compact star surface, and as such offer an
unprecedented opportunity to confront theoretical models of
neutron star surface emission with observations.

The XDINSs X-ray spectrum is with no exception blackbody-like with
temperatures in the range $\sim 40$--100~eV and, thus far,
pulsations have been detected in five sources, with periods in the
range 3--11~s (see Table~\ref{tableins} and refs. therein). In
each of the five cases the pulsed fraction is relatively large
($\sim 12\%$--$35\%$). Quite surprisingly, and contrary to what
one would expect in a simple dipolar geometry, often the hardness
ratio is minimum at the pulse maximum (\citealt{cro2001};
\citealt{hab2003}). Broad absorption features have been detected
around $\sim 300$--700 eV in all pulsating XDINSs and the line
strength appears to vary with the pulse phase. In addition, the
X-ray light curves exhibit a certain asymmetry, with marked
deviations from a pure sinusoidal shape at least in the case of
RBS~1223 \citep[][]{hab2003,schwop05}.
XDINSs were unanimously believed to be steady sources, as indicated by several
years of observations for the brightest of them. Unexpectedly, and
for the first ever time, {\it XMM-Newton\/} observations have recently revealed a
substantial change in the spectral shape and pulse profile of the second most luminous source,
\zeroseven,  over a timescale of $\sim 2$~yr (\citealt{devries04}; \citealt{vink04}). Possible
variations in the pulse profile of  \zerofour \ over a similar timescale ($\sim 0.5$~yr)
have also been reported, although only at a low significance level (\citealt{hanoi2004}).

In the standard picture, emission from an isolated, cooling NS arises when
thermal radiation originating in the outermost surface layers traverses
the atmosphere which covers the star crust. Although the emerging
spectrum is thermal, it is not a blackbody because of radiative transfer
in the magnetized atmosphere and the inhomogeneous surface temperature distribution. The latter
is controlled by the crustal magnetic field, since thermal conductivity across the field
is highly suppressed, and bears the imprint of the field topology. Besides the spectrum,
radiative transfer and the surface temperature distribution act together in shaping the
X-ray lightcurve. Pulse profiles produced by the thermal surface distribution
induced by a simple core-centered dipolar magnetic field have been investigated long
ago by \cite{page95}, under the assumption that each surface patch emits (isotropic)
blackbody radiation. Because of gravitational effects and of the smooth temperature
distribution (the temperature monotonically decreases from the poles to the equator),
the pulse modulation is quite modest (pulsed fraction $\la 10\%$) for reasonable values
of the star radius. Moreover, being the
temperature distribution symmetric about the magnetic equator, the pulse shape itself is
always symmetrical,
regardless of the viewing geometry. Larger pulsed fractions may be reached by
the proper inclusion of an atmosphere. In fact, in a strongly magnetized medium photon
propagation is anysotropic and occurs preferentially along the field (magnetic
beaming, e.g. \citealt{pav94}). Nevertheless, retaining a dipolar temperature distribution
will always result in a symmetric pulse profile.

The quite large pulsed fraction, pulse asymmetry, and possibly long-term variations,
recently observed in XDINSs seem therefore difficult to
explain by assuming that the thermal emission originates at the NS surface,
at least when assuming that the thermal surface distribution is that
induced by a simple core-centered dipolar magnetic field. It should be
stressed that, although the dipole field is a convenient approximation,
the real structure of NSs magnetic field is far from been understood, e.g.
it is still unclear if the field threads the entire star or
is confined in the crust only (e.g. \citealt{gkp04} and references
therein). Whatever the case, there are both observational and theoretical
indications that the NS surface field is ``patchy'' (e.g. \citealt{gep03};
\citealt{urgil04} and references therein). The effects of
a more complex field geometry have been investigated by \cite{pasar96}, who
considered a star-centered dipole+quadrupole field, again assuming isotropic
emission. The presence of multipolar components induces
large temperature variations even between nearby regions and this results in larger
pulsed fractions and asymmetric pulse profiles.

The high quality data now available for thermally emitting NSs, and
XDINSs in particular, demand for a detailed modelling of surface emission
to be exploited to a full extent. Such a treatment should combine both an
accurate formulation of radiation transport in the magnetized atmosphere and
a quite general description of the thermal and magnetic surface distributions,
which, necessary, must go beyond the simple dipole approximation. The ultimate
goal is to produce a completely self-consistent model, capable to reproduce
simultaneously both the spectral and timing properties.
In this paper we take a first step in this direction and present a systematic study
of X-ray lightcurves from cooling NSs, accounting for both a quadrupolar magnetic
field (in addition to the core-centered dipole) and radiative transfer in the magnetized
atmosphere. We computed over 78000 model lightcurves, exploring the entire
parameter space, both in the geometrical angles and the quadrupolar
components. This large dataset has been analyzed using multivariate
statistical methods (the principal component analysis and the cluster
analysis) and we show that a non-vanishing quadrupolar field is required
to reproduce the observed XDINS pulse profiles.


\section{The Model}
\label{model}

\subsection{Going Quadrupolar}
\label{model_quad}

In this section we describe the approach we use to compute the
phase-dependent spectrum emitted by a cooling NS as seen by a
distant observer. This issue has been addressed by several authors
in the past under different assumptions, and basically divides
into two steps. The first involves the computation of the local
(i.e. evaluated by an observer at the star surface) spectrum
emitted by each patch of the star surface while the second
requires to collect the contributions of surface elements which
are ``in view'' at different rotation phases, making proper
account for the fact that only rays propagating parallel to the
line-of-sight (LOS) actually reach the distant observer. Details
on each are presented in the following two subsections
(\S\S~\ref{model_spectra}, \ref{model_tracing}) and further in
Appendix \ref{app1}; here we discuss some general assumptions
which are at the basis of our model.

We take the neutron star to be spherical (mass $M$, radius $R$) and rotating with
constant angular velocity $\omega = 2\pi/P$, where $P$ is the period. Since XDINs are
slow rotators ($P\approx 10$~s), we can describe the space-time outside the NS in terms
of the Schwarzschild geometry (see e.g. \citealt{clm04} for a more complete discussion about
the effects of rotation).

The star magnetic field is assumed to possess a core-centered dipole+quadrupole topology,
$\mathbf B=\mathbf B_{dip}+  \mathbf B_{quad}$. Introducing a polar coordinate
system whose axis coincides with the dipole axis, the (polar) components of
the dipole field at the star surface are

\begin{eqnarray}\label{dipole}
B_{dip,r} &=& f_{dip}B_p\cos\theta\\
\noalign{\smallskip}
B_{dip,\theta} &=& g_{dip}B_p\sin\theta/2\\
\noalign{\smallskip}
B_{dip,\phi} &=&0,
\end{eqnarray}
where $B_p$ is the field strength at the magnetic pole, $\theta$ and $\phi$ are the
magnetic colatitude and azimuth. The functions $f_{dip}$ and $g_{dip}$ account for
the effect of gravity and depend on the dimensionless star radius $x\equiv R/R_S$
with $R_S=2GM/c^2$; their explicit expressions can be found in \citealt[(][see also references therein)]{pasar96}.
The quadrupolar field can be easily derived from the
spherical harmonics expansion and its expression, again at $r=R$,
can be cast in the form

\begin{equation}\label{quadrupole} \mathbf B_{quad} =
\sum_{i=0}^4q_i\mathbf B_{quad}^{(i)} \end{equation} where the $q_i$'s are
arbitrary constants. The polar components of the five generating vectors
$\mathbf B_{quad}^{(i)}$ are reported in \cite{pasar96}. We just note that
their expression for the radial component of the zeroth vector contains a
misprint and should read $B_{quad,r}^{(0)}=(3\cos^2\theta-1)/2$. General
relativistic effects are included in the quadrupolar field by multiplying
the radial and angular components by the two functions $f_{quad}(x)$ and
$g_{quad}(x)$ respectively (see again \citealt{pasar96} for their
expressions and further details).

\begin{figure*} \includegraphics[width=6in,angle=0]{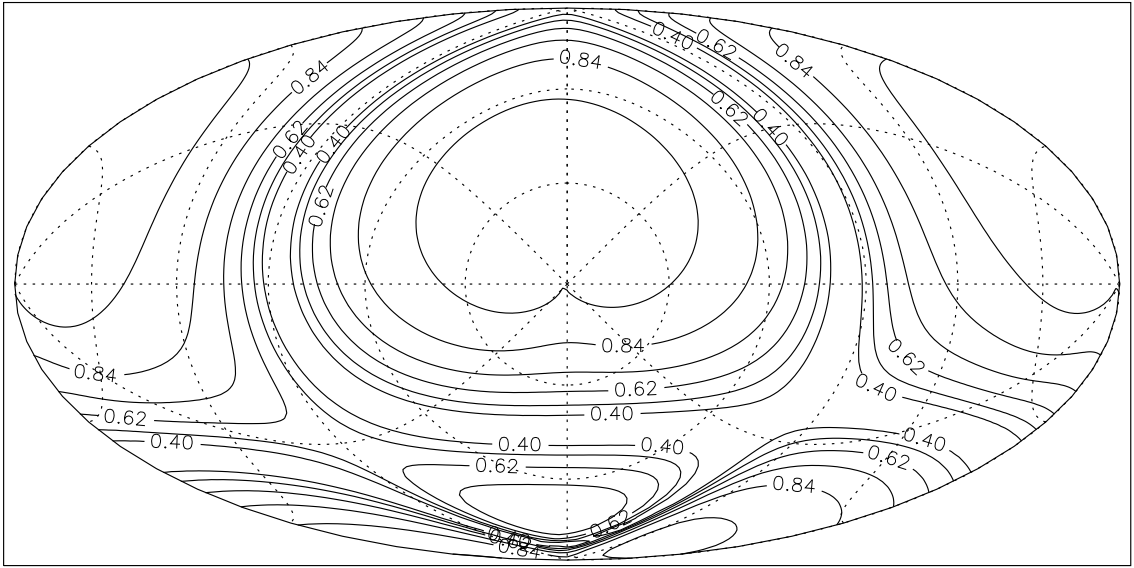}
\caption{\label{map} Hammer projection illustrating the thermal surface distribution
of a NS; the quadrupolar field components (in units of the polar dipolar field) are $b_0=0.0248$,
$b_1=0.684$, $b_2=0.0501$, $b_3=0.0480$, $b_4=-0.0624$. The contours are labelled by
the values of $T_s/T_p$.}
\end{figure*}

The NS surface temperature distribution, $T_s$,  will in general depend on
how heat is
transported through the star envelope. Under the assumption that the field does not
change much over the envelope scale-height, heat transport can be treated
(locally) as
one-dimensional. The surface temperature then depends only
on the angle between the field and the radial direction,
$\cos\alpha=\mathbf B\cdot\mathbf n$, and on the local field strength
$B$ (see \citealt{page95}).
As shown by \cite{grehar83}, an useful approximation is to write

\begin{equation}\label{tsurfgen} T_s =
T_p\left(\cos^2\alpha+\frac{K_\perp}{K_\|}\sin^2\alpha\right)^{1/4}
\end{equation} where the ratio of the conductivities perpendicular
($K_\perp$) and
parallel ($K_\|$)to the field is assumed to be constant. The polar value
$T_p$ fixes the absolute scale of the temperature and is a
model parameter
(\citealt{page95} and references therein). For field strengths $\gg
10^{11}$~G, the conductivity ratio is much less than unity and eq.
(\ref{tsurfgen}) simplifies to

\begin{equation}\label{tsurf} T_s = T_p\vert\cos\alpha\vert^{1/2}\, .
\end{equation} This expression is used in the present investigation.
An example of the thermal surface distribution induced by a
quadrupolar field is shown in figure \ref{map}. Different
approaches which account for the variation of the conductivities
(e.g. \citealt{hh98}) yield similar results. Quite recently
\cite{gkp04} investigated the influence of different magnetic
field configurations on the surface temperature distribution. They
find that, contrary to star-centered core fields, crustal fields
may produce steeper surface temperature gradients. The inclusion
of this effect is beyond the purpose of this first paper. However,
we caveat that, for temperatures expected in XDINs ($\approx
10^6$~K), the differences between the two magnetic configurations
start to be significant when the field strength is $>10^{13}$~G.


\subsection{Radiative Transfer}
\label{model_spectra}

The properties of the radiation spectrum emitted by highly magnetized,
cooling NSs have been thoroughly discussed in the literature (e.g.
\citealt{shi92}; \citealt{pav94}; \citealt{sil01}; \citealt{holai01};
\citealt{holai03}). Since the pressure scale-height is much smaller than
the star radius, model atmospheres are usually computed in the plane
parallel approximation.  Besides surface gravity, the spectrum emerging
from each plane parallel slab depends depends both on the surface
temperature $T_s$ and magnetic field $\mathbf{B}$, either its strength and
orientation with respect to the local normal, which coincides with the
unit vector in the radial direction $\mathbf{n}$.  In order to proceed we
introduce a $(\theta,\, \phi)$ mesh which naturally divides the star
surface into a given number of patches. Once the magnetic field has been
specified, each surface element is characterized by a precise value of
$B$, $\alpha$ and $T_s$. The atmospheric structure and radiative transfer
can be then computed locally by approximating each atmospheric patch with
a plane parallel slab, infinitely extended in the transverse direction and
emitting a total flux $\sigma T_s^4$.

Radiative transfer is solved numerically using the approach described in \cite{don04}.
The code is based on the normal mode approximation for the radiation field propagating
in a strongly magnetized medium and incorporates all relevant radiative processes.
The full angle and energy dependence
in both the plasma opacities and the radiation intensity is retained. In this respect
we note that for an oblique field [i.e. $(\mathbf{B}/B)\cdot\mathbf{n}\neq 1$]
the intensity is not symmetric around $\mathbf{n}$ and depends explicitly on both
propagation angles. If $\mathbf{k}$ is a unit vector along the photon direction,
at depth $\tau$ in the atmosphere the intensity has the form
$I=I_E(\tau,\mu,\varphi)$ where $E$ is the photon energy,
$\mu=\mathbf{n}\cdot\mathbf{k}$ and $\varphi$ is the associated azimuth. Calculations
are restricted to a completely ionized H atmosphere (see \citealt{hoetal03} and
\citealt{pocha03} for a treatment of ionization in H atmospheres).

Since the numerical evaluation of model atmospheres is computationally quite
demanding, especially for relatively high, oblique fields and $T_s < 10^6$~K, we
preferred to create an archive beforehand, by computing models for preassigned values
of $\cos\alpha$, $B$ and $T_s$. The range of the latter
two parameters should be wide enough to cover the surface variation of $B$ and $T$ in all
the cases of interest: $12\leq\log B\leq 13.5$ and $5.4\leq\log T_s\leq 6.6$.
Particular care was taken to generate models in the entire $\alpha$ domain,
$0\leq\cos\alpha\leq 1$. According to the adopted
surface temperature distribution (eq. [\ref{tsurf}]), the regions close to the
equator have very low temperatures and can not be associated with any model in
the archive. However, being so cool, their contribution to the observed spectrum is
negligible (see \S\ref{model_numerics}). For each model the emerging radiation intensity,
$I_E(\tau=0,\mu,\varphi)$, is stored, for $0.01\, {\rm keV}\leq E\leq 10\, {\rm keV}$,
$0\leq\mu\leq 1$ and $0\leq\varphi\leq 2\pi$. The archive itself consists of a six-dimensional
array ${\cal I}(B,T_s,\cos\alpha;E,\mu,\varphi)$ which associates at each set of
the parameters $B,\, \cos\alpha,\, T_s$ the (discrete) values of the angle- and
energy-dependent intensity.
Actually, since the code makes use of adaptive computational meshes, the emerging
intensities have been interpolated on common energy and angular grids before storage.
The final array contains the emerging intensity for about 100
atmospheric models, evaluated at 100 energy bins and on a $20\times 20$
angular mesh, $(\mu,\varphi)$.


\subsection{The Observed Spectrum}
\label{model_tracing}

The problem of computing the pulse profile produced by hot caps
onto the surface of a slowly rotating NS including gravity effects was first
tackled by \cite{pfc83}. Their results were then generalized to the
case of emission from the entire star surface with an assigned temperature
distribution by \cite{page95} and \cite{donros04}, in case of isotropic and
non-isotropic radiation fields respectively. The approach used here follows
that discussed in the two papers quoted above which we refer to for more details.
For the sake of clarity, we will present a Newtonian derivation first.
Relativistic ray-bending can be then accounted for quite straightforwardly.

The NS viewing geometry is described in terms of two angles $\chi$ and $\xi$
which give the inclination of the LOS and of the dipole axis with respect
to the star spin axis. Let $\mathbf z$, ${\mathbf b}_{dip}$ and
${\mathbf p}$ denote the unit vectors along the same three directions.
Let us moreover introduce two cartesian coordinate systems, both with
origin at the star center: the first,
$(X,\, Y,\, Z)$, is fixed and such that the $Z$-axis coincides
with the LOS while the $X$-axis is in the $(\mathbf z,\, \mathbf p)$ plane;
the second, $(x,\, y,\, z)$, rotates with the star. The $z$-axis is parallel to
${\mathbf b}_{dip}$ while the choice of the $x$-axis will be made shortly.
Each cartesian frame has an associated polar coordinate system, with polar axes
along $Z$ and $z$, respectively. The colatitude and azimuth
are $(\Theta,\, \Phi)$ in the fixed frame, and $(\theta,\, \phi)$
in the rotating one (the latter are just the magnetic colatitude and azimuth
introduced in \S~\ref{model_quad}).

In the following we shall express vectors through their components: these are always
the cartesian components referred to the fixed frame, unless otherwise explicitly
stated. The same components are used to evaluate both scalar and vector products.
Upon introducing the phase angle $\gamma=\omega t$, it follows from elementary
geometrical considerations that ${\mathbf p}= (-\sin\chi,0,\cos\chi)$ and
${\mathbf b}_{dip}=(-\sin\chi\cos\xi-\cos\chi\sin\xi\cos\gamma,
-\sin\xi\sin\gamma,\cos\chi\cos\xi+\sin\chi\sin\xi\cos\gamma)\, .$
It can be easily verified that ${\mathbf q}=(\cos\xi\sin\gamma,
\cos\gamma,\sin\xi\cos\gamma)$ represents an unit vector orthogonal to ${\mathbf p}$
and rotating with angular velocity $\omega$. We then choose the $x$-axis in the direction of
the (normalized) vector component of ${\mathbf q}$ perpendicular to ${\mathbf b}_{dip}$,
\begin{equation}
\label{qperp}
{\mathbf q}_\perp = \frac{{\mathbf q}-\left({\mathbf b}_{dip}\cdot{\mathbf q}\right)
{\mathbf b}_{dip}}{\left[1-\left({\mathbf b}_{dip}\cdot{\mathbf q}\right)^2\right]^{1/2}}\, ;
\end{equation}
the $y$-axis is parallel to ${\mathbf b}_{dip}\times {\mathbf q}_\perp$.
The local unit normal relative
to a point on the star surface of coordinates ($\Theta,\, \Phi$) is readily
expressed as ${\mathbf n}=(\sin\Theta\cos\Phi, \sin\Theta\sin\Phi,\cos\Theta)$. By
introducing the unit vector ${\mathbf n}_\perp$, defined in strict analogy with
${\mathbf q}_\perp$ (see eq. [\ref{qperp}]), the two expressions

\begin{eqnarray}
\label{theta}
\cos\theta &=& {\mathbf b}_{dip}\cdot{\mathbf n} \\
\noalign{\smallskip}
\label{phi}
\cos\phi &=&{\mathbf n}_\perp\cdot{\mathbf q}_\perp
\end{eqnarray}
provide the relations between the two pairs of polar angles, the geometrical angles
$\xi,\, \chi$ and the phase. While direct inversion of (\ref{theta}) poses no problems since
it is $0\leq\theta\leq\pi$, care must be taken to ensure that $\phi$, as obtained from
(\ref{phi}), covers the entire range $[0,2\pi]$. This is achieved by replacing $\phi$
with $2\pi-\phi$ when ${\mathbf n}\cdot({\mathbf b}_{dip}\times {\mathbf q}_\perp)<0$.

We are now in the position to compute the total monochromatic flux emitted by the star and
received by a distant observer. This is done by integrating the specific intensity
over the visible part of the star surface at any given phase (see e.g. \citealt{donros04})
\begin{equation}\label{fluxint}
F_E(\gamma)\propto \int_0^{2\pi}\, d\Phi\int_0^1 {\cal I}(B,
T_s,\cos\alpha;E, \mu,\varphi)\, du^2
\end{equation}
where $u=\sin\Theta$. Further integration of eq.~(\ref{fluxint}) over $\gamma$ provides
the phase-averaged spectrum. As discussed in
\S~\ref{model_spectra}, the intensity depends on the properties of the surface
patch and on the photon direction. The magnetic field strength $B$ can be directly computed
from the polar components of $\mathbf B$
(see \S~\ref{model_quad} and \citealt{pasar96}).
The magnetic tilt angle $\alpha$ and the surface temperature (see eq. [\ref{tsurf}])
follow from $\cos\alpha={\mathbf n}\cdot{\mathbf B}/B=B_r/B$, being $\mathbf n$ the unit
radial vector. The local values of $\mathbf B$ and $T_s$ depend on ($\theta,\, \phi$). They
can be then easily expressed in terms of $(\Theta,\, \Phi)$ for any given phase using eqs.
(\ref{theta})-(\ref{phi}). Because the star appears point-like, there
is a unique ray which reaches the observer from any given surface element and this implies
that also $\mu$ and $\varphi$ are functions of $(\Theta,\, \Phi)$. It is
clearly
$\mu=\cos\Theta$, while $\varphi$ is given by $\cos\varphi={\mathbf m}\cdot{\mathbf v}$.
The two unit vectors which enter the previous expression are, respectively, the projections
of ${\mathbf B}$ and ${\mathbf z}$ on the plane locally tangent to the surface. They are
expressed as ${\mathbf m}=(\cos\Theta\cos\Phi, \cos\Theta\sin\Phi,-\sin\Theta)$ and
${\mathbf v}=({\mathbf B}/B-{\mathbf n}\cos\alpha)/\sin\alpha$.
The cartesian components of ${\mathbf B}$ needed to evaluate $\cos\varphi$ are derived in
the Appendix.

Gravity effects (i.e. relativistic ray-bending) can be now included in a very simple way.
The local value of the colatitude $\Theta$ is, in fact, related to that
measured by an observer at infinity by the ``ray-tracing'' integral
\begin{equation}\label{raytrac}
\bar\Theta=\int_0^{1/2}u\left[\frac{1}{4}\left(1-\frac{1}{x}\right)-
\left(1-\frac{2v}{x}\right)x^2v^2\right]^{-1/2}dv
\end{equation}
where $x=R/R_S$.
Since we are collecting the contributions of all surface elements seen by a distant
observer, each patch is labelled by the two angles $\bar\Theta$ and $\Phi$. This means
that the integrand in (\ref{fluxint}) is to be evaluated precisely at the same
two angles and is tantamount to replace $\Theta$ with $\bar\Theta$ in all previous
expressions. Note, however, that the innermost integral in (\ref{fluxint}) is always
computed over $\Theta$.

\begin{figure*}
\includegraphics[height=6.truecm,width=8.truecm, angle=0]{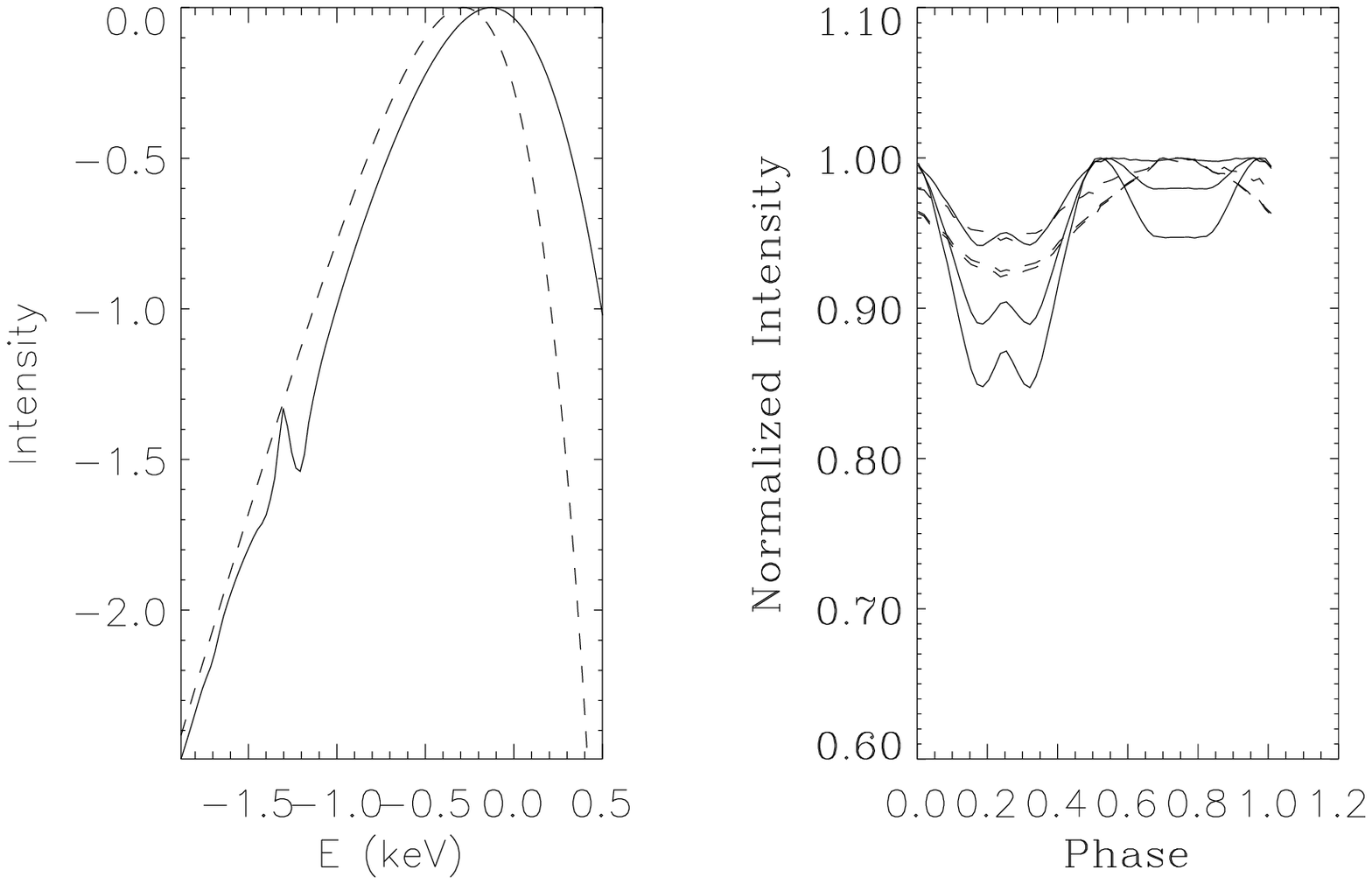}
\includegraphics[height=6.truecm,width=8.truecm, angle=0]{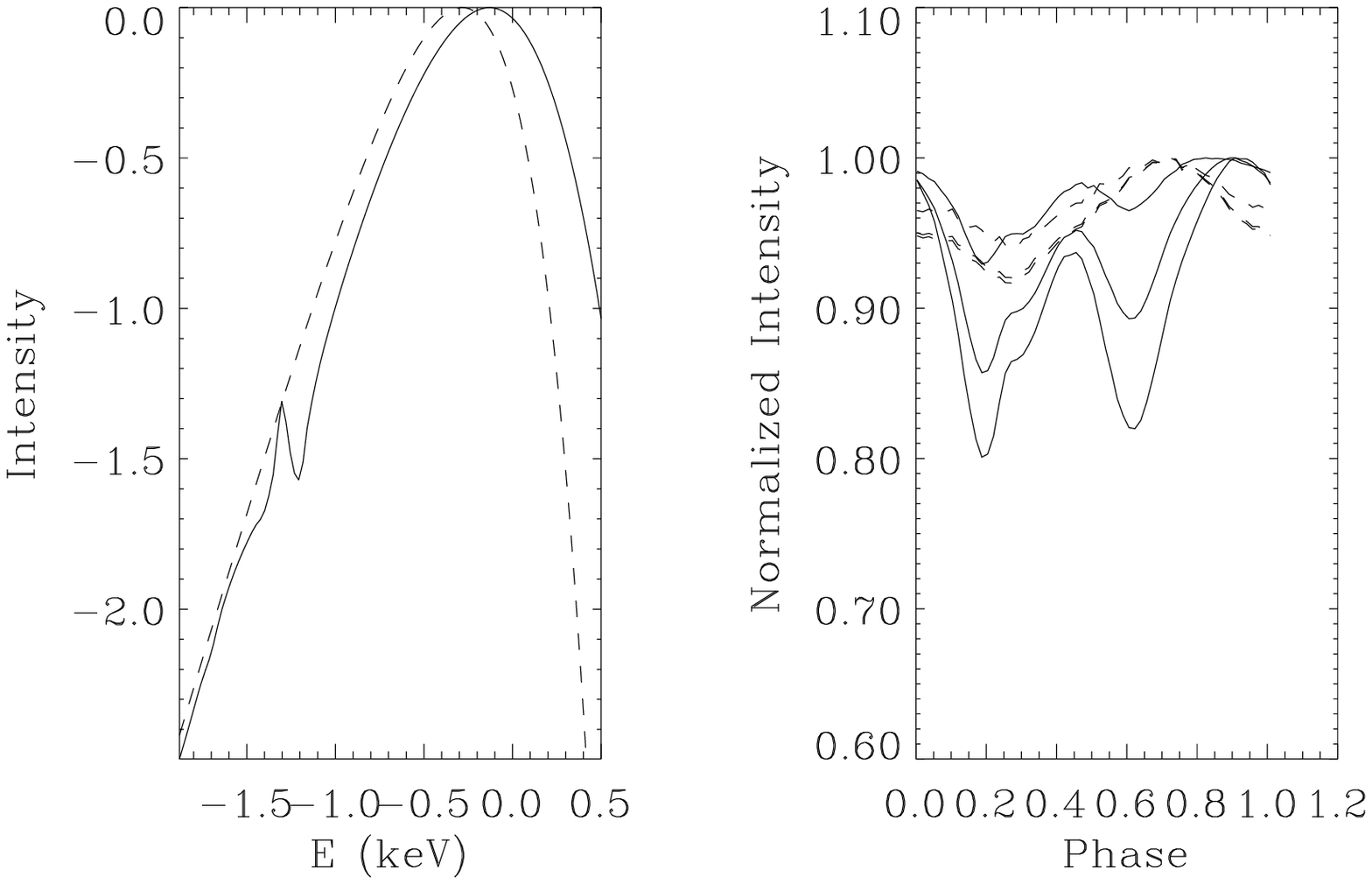}
\includegraphics[height=6.truecm,width=8.truecm, angle=0]{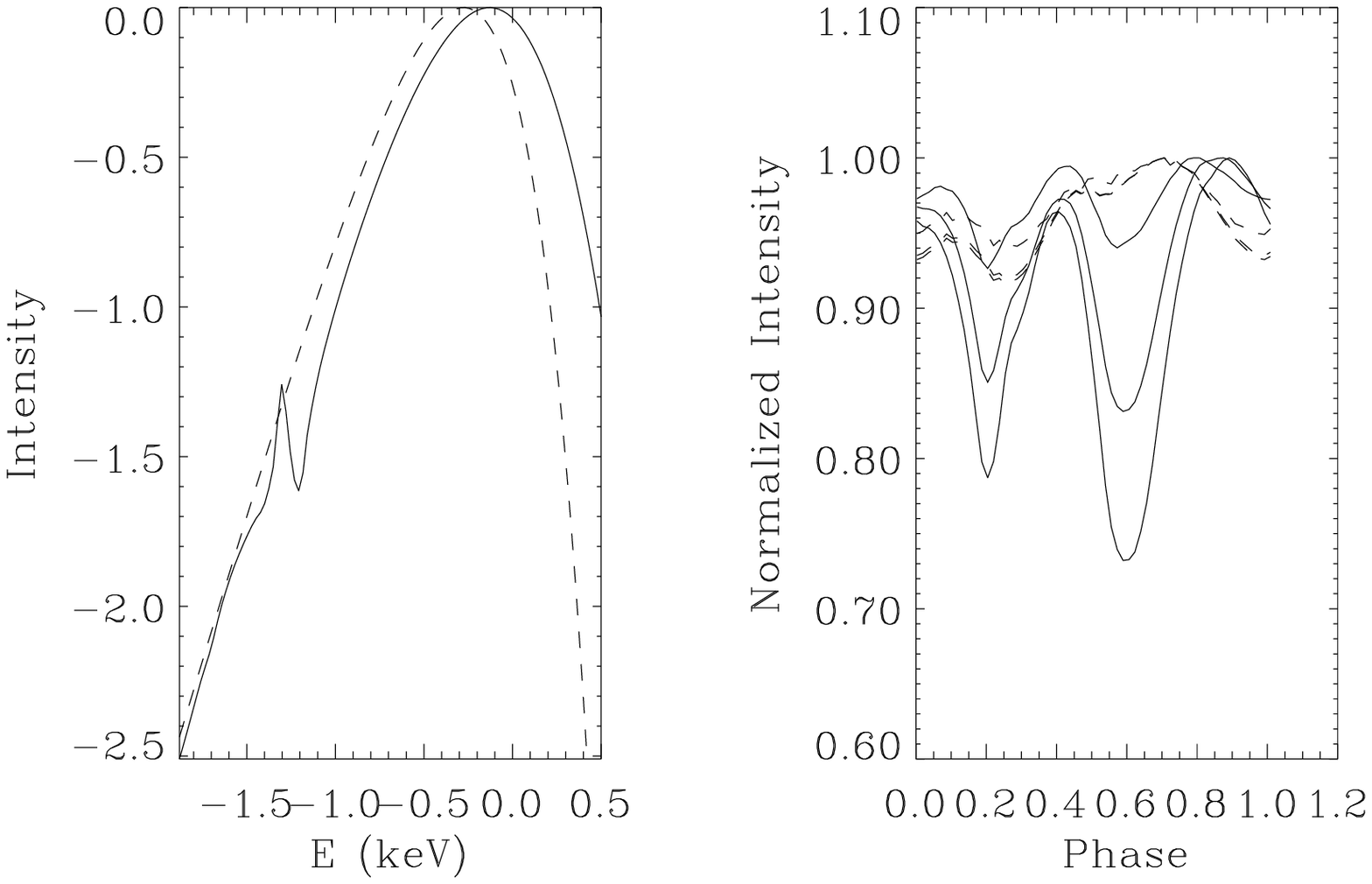}
\includegraphics[height=6.truecm,width=8.truecm, angle=0]{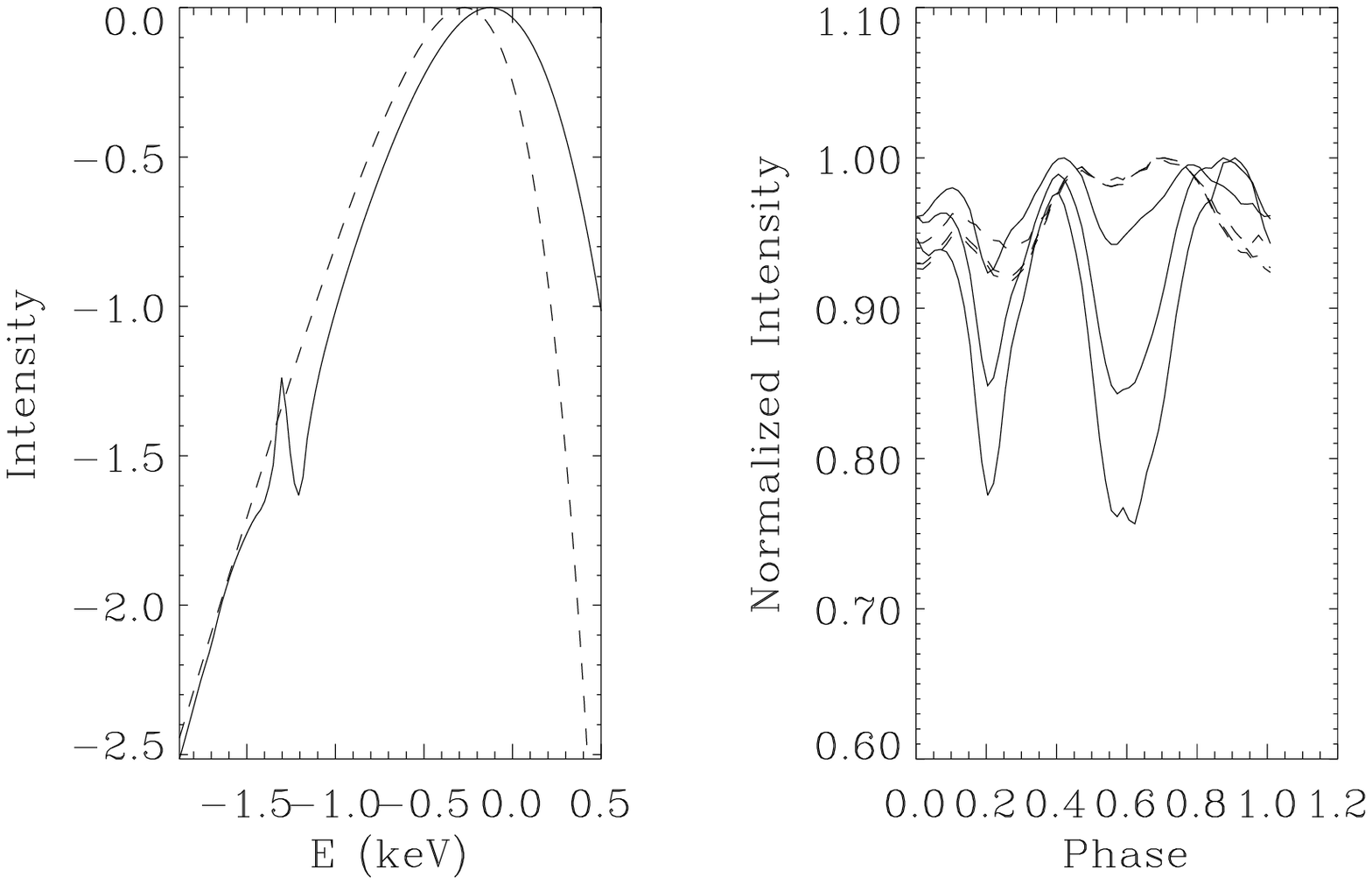}
\caption{\label{beam} Effects of Radiative beaming. In each pair of 
figures we show
phase average spectrum (left) and light curves in the 0.1-2~keV, 0.1-0.5~keV and
0.5-2~keV bands (right, arbitrary normalization). In all panels solid 
lines refer to atmospheric models, while dashed lines to blackbody 
emission.
The four pairs of figures correspond to different values of $\xi$: from top left to
bottom right it is $\xi=0^\circ, 30^\circ, 60^\circ, 90^\circ$, respectively. The
remaining parameters are fixed at: $B_{dip} = 6x10^{12}$~G, $T_{pol}=2.5$MK, $b_0=0.5$,
$b_1=b_3=b_4=0$, $b_2 = 0.9$, $\chi=90^\circ$.} \end{figure*}

Effects of radiative beaming are illustrated in fig.~\ref{beam}, were we
compare phase average spectra and lightcurves computed by using radiative
atmosphere model with those obtained under the assumption of isotropic
blackbody emission. As we can see, the pulse profiles are substantially
different in the two cases. Also, by accounting for radiative beaming
allow to reach relatively large pulse fractions ($\sim 20$\%).

\subsection{Numerical Implementation}
\label{model_numerics}

The numerical evaluation of the phase-dependent spectrum has been carried out
using an IDL script. Since our immediate goal is to check if the observed pulse
profiles can be reproduced by our model, we start by computing the lightcurve in
a given energy band, accounting for interstellar absorption and the detector response
function. Most of the recent observations of X-ray emitting INSs have been obtained by
{\it XMM-Newton\/}, so here we refer to EPIC-pn response function.
Both absorption and the detector response depend upon the arrival photon energy
$\bar E=E\sqrt{1-1/x}$, so the pulse profile in the $[\bar E_1,\, \bar E_2]$ range
is given by
\begin{eqnarray}\label{lcband}
F(\gamma)&\propto&
\int_0^{2\pi}\, d\Phi\int_0^1du^2\int_{\bar E_1}^{\bar E_2} A(\bar E)
\exp{[-N_H\sigma(\bar E)]}\nonumber\\
&&\times {\cal I}(B, T_s,\cos\alpha;E, \mu,\varphi)\, d\bar E \\
& \equiv & \int_0^{2\pi}\, d\Phi\int_0^1du^2 {\cal J} \nonumber
\end{eqnarray}
where $A$ is the response function, $N_H$ is the column density, and $\sigma$ the
interstellar absorption cross section (e.g. \citealt{mormc83}).

Since the energy integral $\cal J$ does not involve geometry, it is
evaluated first. We select
three energy bands, corresponding to the soft (0.1--0.5~keV) and
hard (0.5--1~keV) X-ray colors, and to the total range 0.1--5~keV. Results are
stored as described in \S~\ref{model_spectra} in the case of the quantity
${\cal I}$, the only difference
being that energy index in the array $\cal J$ runs now from 1 to 3 in
correspondence
to the three energy intervals defined above. We then introduce a
$(\mu=\cos\Theta,\, \Phi)$
mesh by specifying $50\times 50$ equally spaced points in the $[0,\, 1]$ and  $[0,\, 2\pi]$
intervals, and interpolate the intensity array at the required values of $\mu$, $\mu=u$.
Next, the values of $\bar\Theta$ relative to the $u$ grid are computed from eq. (\ref{raytrac}).
All these steps can be performed in advance and once for all, because
they do not depend on the viewing geometry or the magnetic field.

Then, once the angles $\chi$, $\xi$ and the phase $\gamma$ have been
specified, the magnetic
colatitude and azimuth, $\theta(\bar\Theta,\Phi,\gamma)$, $\phi(\bar\Theta,\Phi,\gamma)$ can be
evaluated. We use 32 phase bins and a set of five values for each of
the two angles $\chi$, $\xi=(0^\circ,30^\circ,50^\circ,70^\circ,90^\circ)$.
The magnetic field is assigned by prescribing the strength of the quadrupolar components relative
to polar dipole field, $b_i=q_i/B_p$ $(i=1,\ldots, 5)$, in addition to
$B_p$ itself; in our grid, each
$b_i$ can take the values $(0,\pm 0.25,\pm 0.5)$. For each pair $\theta,\phi$ we then
compute $\mathbf B$ and $\cos\varphi$; $\cos\alpha$ gives the surface temperature $T_s$ once
$T_p$ has been chosen. The corresponding values of the intensity are
obtained by linear interpolation
of the array $\cal J$. Surface elements emitting a flux two order of
magnitudes lower than that of the polar region ($\sigma T_p^4$) were assumed
not to give any contribution to the observed spectrum.
Finally, direct numerical evaluation of the two angular integrals in (\ref{lcband}) gives
the lightcurves. Although in view of future application we computed and
stored the lightcurves in the
three energy bands mentioned above, results presented in \S~\ref{stat} and
\ref{obs} are obtained using always the total (0.1-5~keV) energy band.
To summarize, each model
lightcurve depends on
$\chi$, $\xi$, and the five $b_i$. No attempt has been made here to vary also $T_p$ and $B_p$,
which have been fixed at $140$~eV and $6\times 10^{12}$~G respectively. A total of 78125 models
have been computed and stored. Their analysis is discussed in \S~\ref{stat}.


\section{Analyzing lightcurves as a population}
\label{stat}

As discussed in \S~\ref{model}, under our assumptions the computed
lightcurve is a multidimensional function which depends in a complex way
on several parameters. Therefore, an obvious question is whether or not we
can identify some possible combinations of the independent parameters that
are associated to particular features observed in the pulse shape. The
problem to quantify the degree of variance of a sample of individuals (in
our case the lightcurves) and to identify groups of ``similar'' individuals
within a population is a common one in behavioral and social
sciences. Several techniques have been extensively detailed in many books
of multivariate statistics (e.g.  \citealt{kendall1957};
\citealt{manly1998})  and, although they have been little used in physical
sciences, a few applications to astrophysical problems have been presented
over the past decades (see e.g. \citealt{whitney1983}; \citealt{mit90};
\citealt{hey97}).

We focus here on a particular tool called {\it principal
components analysis} (PCA), which appears promising for a
quantitative classification of the lightcurve features. The main
goal of PCA is to reduce the number of variables that need to be
considered in order to describe the data set, by introducing a new
set of variables $z_p$ (called the principal components, PCs)
which can discriminate most effectively among the individuals in
the sample, i.e. the lightcurves in our present case. The PCs are
uncorrelated and mutually-orthogonal linear combinations of the
original variables.
Besides, the indices of the PCs are ordered so that $z_1$ displays the largest
amount of variation, $z_2$ displays the second largest and so on, that is,
$\mathrm{var}(z_1) \geq \mathrm{var}(z_2) \geq\ldots \geq
\mathrm{var}(z_p)$ where $\mathrm{var}(z_k)$ is the variance in the sample
associated with the $k$-th PC.
Although the physical meaning of the new variables may be in general not immediate,
it often turns out that a good representation of the population is
obtained by using a limited number of PCs, which allows to treat the data
in terms of their true dimensionality.
Beyond that, PCA represents a first step towards other kind of multivariate
analyses, like the  {\it cluster analysis}.  This is a
tool which allows to identify subgroups of objects so that ``similar''
ones belong to the same group.  When applying the cluster analysis
algorithm a PCA is performed first in order to reduce the number of original
variables down to a smaller number of PCs. This can substantially
reduce the computational time.

Since both tools are extensively described in the literature,
we omit all mathematical details
for which an interested reader may refer to, e.g.,
\cite{kendall1957} and \cite{manly1998}. Let us denote with
$y_{ps}$ $(p=1,\ldots,\, P;\ s=1, \ldots,\, S)$ the values of the
intensity computed at each phase bin, for each model lightcurve. Let us also
introduce the ``centered'' data $x_{ps}$, defined as
\begin{equation}
x_{ps}  = \left ( y_{ps} - \mu_p \right) / s_p  \, ,
\end{equation}
where
$\mu_p$ and $s_p$ are the mean and the variance of the computed data, respectively.
In order to shift to the new PCs variables $z_{ps}$, we computed the
transformation matrix $V'$
such that
\begin{equation}
\label{pcrepre}
z_{ps}  = \sum_q v'_{pq} x_{qs}  \, .
\end{equation}
A sufficient condition to specify univocally $V'$ is to impose that
the axes of the new coordinate system (i.e. the PCs) are mutually
orthogonal and linearly independent.

By applying the PCA to our sample of models, we have found that
each lightcurve can be reproduced using only the first ~20-21 more
significant PCs (instead of 32 phases) and that the first 4 PCs
alone account for as much as $\sim 85\%$ of the total variance.
Moreover, $\sim 72\% $
of the variance is in the first three PCs only. It is therefore meaningful to
introduce a graphical
representation of the model dataset in terms of the first three $z_i$'s.
This is shown in figure \ref{pca} where black/red squares gives the
position in the $z_1z_2z_3$ space of quadrupolar/dipolar models.
To better
visualize the latter, an additional set of lightcurves was
computed, bringing the total number of dipolar models displayed in
fig. \ref{pca} to 100.

\begin{figure*} \includegraphics[width=6in,angle=0]{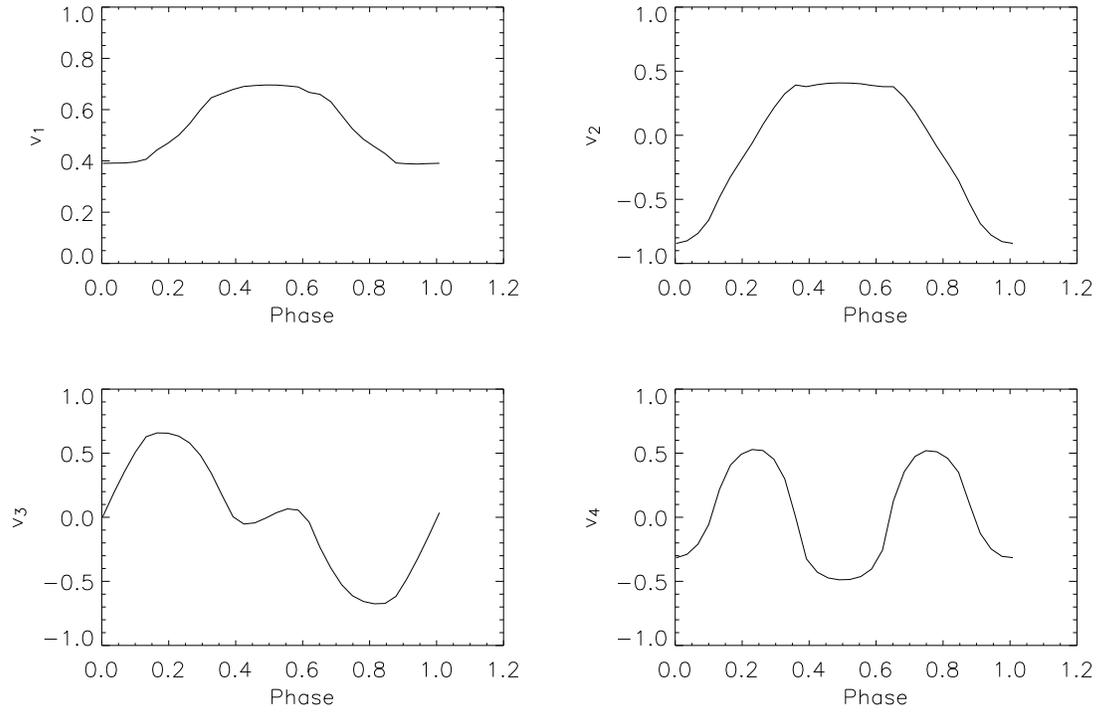}
\caption{\label{vp}The first four coefficients $v_p$ ($p=1...4$)
with respect to the phase $\gamma$. See text for all details.
}
\end{figure*}

An insight on the lightcurve property each of the PCs measures can
be obtained by inspecting the coefficients $v'_{pq} $ of the linear
combination which gives the $z_p$'s for an assigned dataset
[see eq. (\ref{pcrepre})]. Since $z_{p}  = \sum_q v'_{pq} x_{q}\propto
\int_0^{2\pi} v_p(\gamma)F(\gamma)\, d\gamma$, this is tantamount to
assess the effect of the filter $v_p(\gamma)$ on the lightcurve $F(\gamma)$
knowing the values of the former at a discrete set of phases. The first
four $v_p$ are shown in Fig.~\ref{vp}. The first PC provides a measure of
the amplitude of the
pulse; it is always
$z_1>0$, and
large values of $z_1$ correspond to low pulsed fractions. Both $z_2$ and $z_3$
may take either sign (same for higher order PCs) and give information on
the pulse shape.
We note that, although the
absolute phase is in principle an arbitrary quantity, the whole models
population has been computed using the same value. Therefore, when
studying the morphological properties of the sample of lightcurves, it is
meaningful to refer to the symmetry properties with respect to
the parameter $\gamma$.
Large and negative values of $z_2$ imply that the main
contributions to the lightcurve comes from phases around zero. $z_3$ measures
the parity of the lightcurve with respect to half period. Pulses which are
symmetric have $z_3=0$. As fig. \ref{pca} clearly shows, the PCA is very effective
in discriminating purely dipolar from quadrupolar models. The former cluster
in the ``tip of the cloud'', at large values of $z_1$, negative values of $z_2$ in
the $z_3=0$ plane, as expected. In fact, dipolar pulse patterns are always
symmetrical and their pulsed fraction is quite small (semi-amplitude
$\la 10\%$). It is worth noticing that quadrupolar magnetic configurations too
can produce quite symmetrical lightcurves, e.g. the black squares in fig. \ref{pca}
with $z_3\sim 0$. However, in this case the pulsed fraction may be much larger, as
indicated by the lower values of $z_1$ attained by quadrupolar models with $z_3=0$
in comparison with purely dipolar ones. This implies that a symmetrical lightcurve
is not {\it per se\/} an indicator of a dipolar magnetic geometry.

As suggested by some authors (e.g.
\citealt{heck76}; \citealt{whitney1983}), PCs can then be used as
new variables to describe the original data. However, in the case
at hand, the problem is that  although PCs effectively distinguish
among pulse patterns, they have a purely geometrical meaning
and can not be directly related to the physical parameters
of the model ($b_i, \chi, \xi$). We have found that the standard regression
method does not allow to link the PCs with the model parameters, which
is likely to be a signature of a strong non-linearity (see
\S~\ref{discuss}).
Instead, the PCA can be regarded as a method to provide a link between the
numerous lightcurves, in the sense that models ``close'' to each other in the
PCs space will have similar characteristics. Unfortunately, despite
different definitions of metrics have been attempted, so far we found it difficult to
translate the concept of ``proximity'' in the PCs space in a corresponding
``proximity'' in the 7-dimensional space of the physical parameters $\xi$,
$\chi$ and $b_i$ $(i=1,\ldots, 5)$.
By performing a cluster analysis we found
that two separate subgroups are clearly evident in the PCs space, one of
which encompasses the region occupied by purely dipolar models
(see
fig.~\ref{clustfit0420}, top panel). However, again
it is not immediate to find a corresponding subdivision in the physical
space. Due to these difficulties,
we postpone a more detailed statistical analysis
to a follow-up paper, and, as discussed in the next section,
we concentrate on the direct application of our
models to the observed lightcurves of some isolated neutron stars.


\section{An Application to XDINSs}
\label{obs}

\begin{figure*} 
\vskip 2.truecm
\centerline{{\bf This figure is provided separately as the 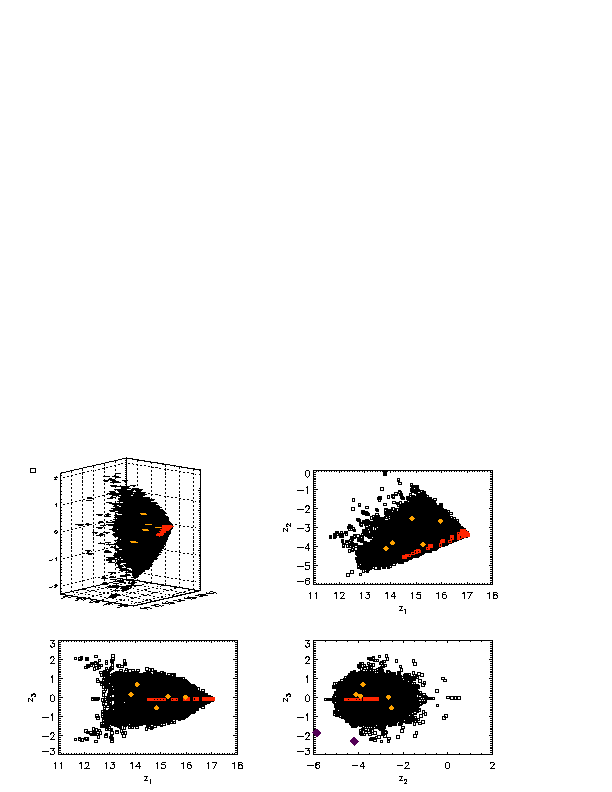 file}}
\vskip 2.truecm
\caption{\label{pca}The computed population of lightcurves plotted
against the first 3 principal components; units on the axes are
arbitrary. Yellow symbols mark the position of the EPIC-PN light
curves of XDINSs (both observations relative to rev.~78 and
rev.~711 are shown for RX J0720.4-3125), red symbols mark the
lightcurves computed assuming a pure dipolar field. The PCs
representation (limited to the first three PCs, which alone
account for the 72\% of the total variance) of the observed
lightcurves falls within the domain spanned by the quadrupolar
model representations; this is why at least one fitting solution
can be found. For comparison, the position of the EPIC-PN
lightcurves of the Anomalous X-ray pulsar 1E~1048.1-5937 is also
shown. These two points fall outside the model distribution
and they can be seen only in the last panel (violet diamonds;
see text for details).}
\end{figure*}

\begin{figure}
\vskip 3.truecm
\centerline{{\bf The top panel of fig.5 is provided 
separately}}
\centerline{{\bf as the 
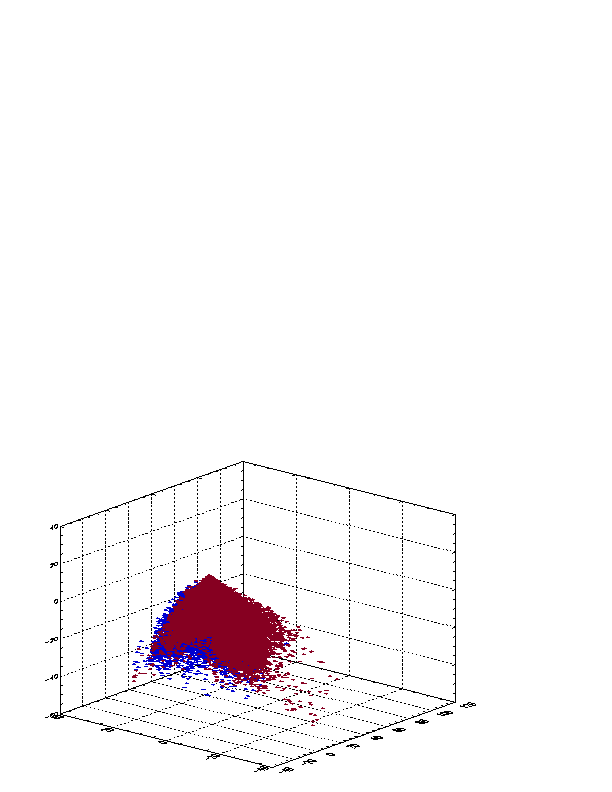 file}}
\null \vskip 3.truecm
\hfil\hspace{\fill}
\begin{minipage}{95mm}
\includegraphics[width=90mm,angle=0]{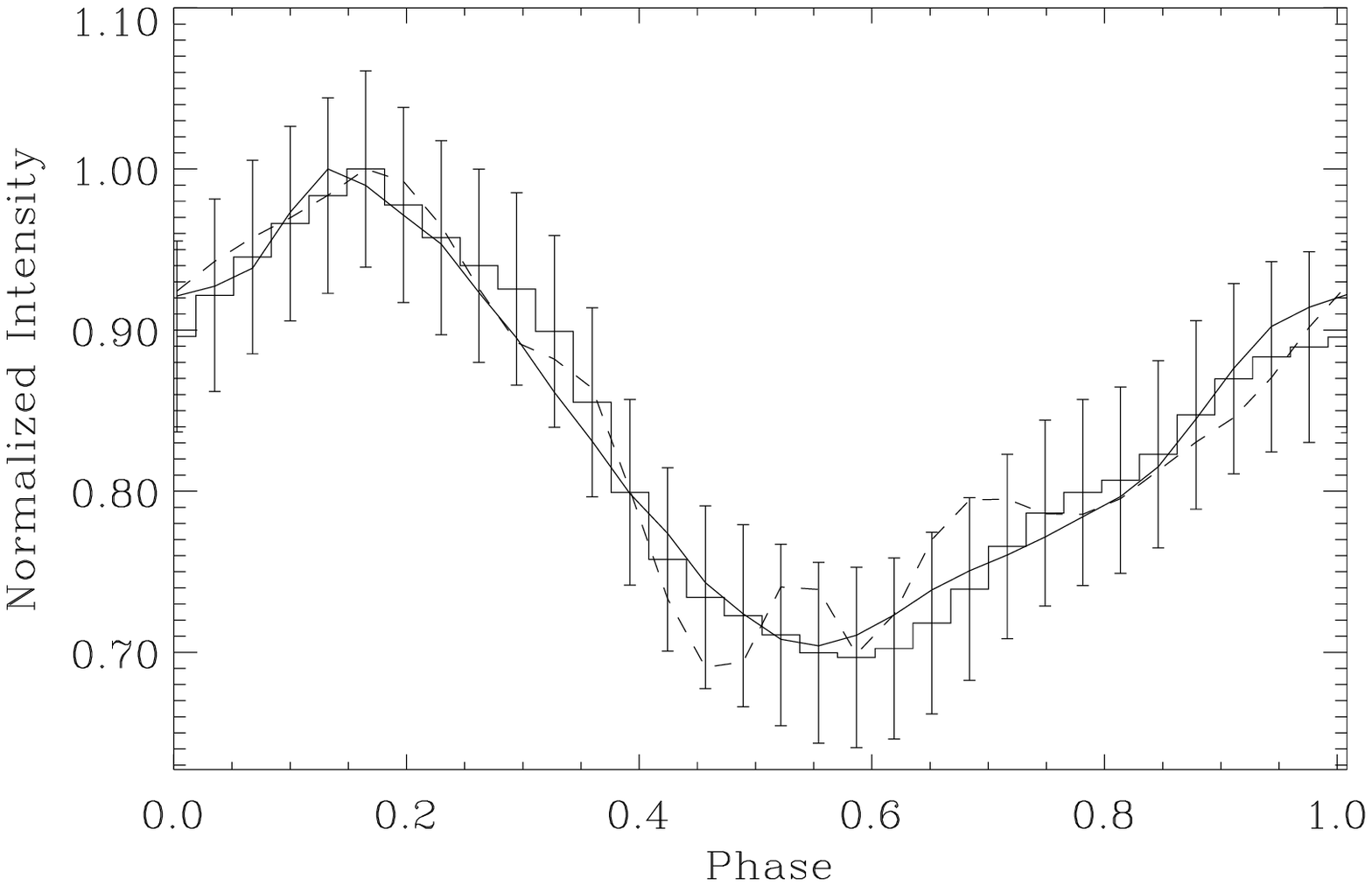}
\end{minipage}
\caption{\label{clustfit0420} Top: Results of a cluster analysis
performed on the models population. As in fig.~\ref{pca}, axis are
the first three PCs: the different scale is due to the fact that
here the PCs have been computed on the centered data.
Two separate subgroups, plotted in red and
blue respectively, are evident in the PCs space. Bottom: Fit of
the EPIC-PN (0.12-0.7 keV) lightcurve of RX J0420.0-5022 detected
during rev.~570 (\citealt{hanoi2004}). Data point refer to the
smoothed observed lightcurve; dashed line: trial solution as
inferred from the closest model in the PCs space; solid line: best
fit solution. The best fit parameters are: $b_0 = -0.48$,  $b_1 =
0.02$,  $b_2 = -0.25$,  $b_3 = 0.35$,  $b_4 = -0.20$, $\xi =
39.9^\circ$, $\chi=91.2^\circ$.}
\end{figure}

In the light of the discussion at the end of the previous section, the PCA
may be used to get insights on the ability of the present model to
reproduce observed lightcurves. A simple check consists in deriving the PC
representation of the pulse profiles of a number of sources and verify if
the corresponding points in the $z_1z_2z_3$ space fall inside the volume
covered by the models. We stress
once again that, although the model lightcurves depend upon all the PCs,
$z_1$, $z_2$, and $z_3$ alone provide a quite accurate description of the
dataset since they account for a large fraction of the variance. In this
sense the plots in fig. \ref{pca} give a faithful representation of the
lightcurves, with no substantial loss of information: profiles close to
each other in this 3-D space exhibit a similar shape. To this end, we took
the published data for the first four
pulsating XDINSs listed in table \ref{tableins} and rebinned the
lightcurves
to the same 32 phase intervals used for the models. Since the PCA of the
model dataset already provided the matrix of coefficients $v'_{pq}$ (see
eq. [\ref{pcrepre}]), the PCs of any given observed lightcurve are
$z_{p}^{obs} = \sum_q v'_{pq} x_{q}^{obs}$, where $x_{q}^{obs}$ is the
(standardized) X-ray intensity at phase $\gamma_q$. As it can be seen from
fig. \ref{pca}, the observed pulse profiles
never fall in the much smaller region occupied by purely dipolar models.
However, they all lie close to the quadrupolar
models, indicating that a quadrupolar configuration able to reproduce the
observed features may exist.
A possible exception to this is the first observation of RX J0720.4-3125
({\it XMM-Newton\/} rev. 078), for which the pulse profile appears quite
symmetric and the pulsed fraction is relatively small (see table
\ref{tableins}). While a purely dipolar configuration may be able to
reproduce the lightcurve for rev. 078, a visual inspection of fig.
\ref{pca} shows that this is not the case for the second observation of
the same source ({\it XMM-Newton\/} rev. 711). Despite the pulse shape is
still quite symmetrical, the pulsed fraction is definitely too large to be
explained, within the current model, with a simple dipole field. The same
considerations apply to the lightcurve of RX J0420.0-5022.

Just as a counter-example, we added in fig. \ref{pca} the PC
representation of the X-ray lightcurve of the Anomalous X-ray
pulsar 1E~1048.1-5937 observed in two different epochs, 2000
December and 2003 June (see \citealt{sandro2004}). The new data
points fall well outside the region of the populated by
quadrupolar models and can be seen only in the last panel of
fig.~\ref{pca}. In the case of this source, the pulsed fraction is
so high ($89\%$ and $53\%$ in June and December respectively) that
no quadrupolar configuration could account for it. In terms of
PCs, both points have a relatively low value of $z_1$ ($z_1=9.9$
and $z_1=6.9$). In this case no fit can be found, not
surprisingly, since we do expect a large contribution from a
non-thermal power law component to the X-ray emission of Anomalous
X-ray pulsars.

To better understand to which extent quadrupolar surface distributions may
indeed produce lightcurves close to the observed ones, we select the model
in the data set which is ``closer'' to each of the observed lightcurves.
This is done looking first for the minimum Euclidean distance between the
observed and the model pulse profiles in the PCs space. Note that in this
case all the relevant PCs were used, that is to say the distance is
defined by $D^2=\sum_{p=1}^{20} (z_p- z_{p}^{obs})^2$. The computed model
which minimizes $D$ is then used as the starting point for a fit to the
observed lightcurve which yields the best estimate of the model
parameters. The fitting is performed by computing ``on the fly'' the
lightcurve for the current values of the parameters and using the standard
(i.e. not the PC) representation of both model and data. The quadrupolar
components and viewing angles are treated as free parameters while the
polar values $T_p$ and $B_{dip}$ are fixed and must fall in the domain
spanned by our atmosphere model archive.\footnote{In general this will not
contains the exact values inferred from spectral observations of XDINSs.
However, we have checked that a fit (albeit with different values of the
quadrupolar field) is possible for different combinations of $B_{dip}$ and
$T_p$ in the range of interest.} The (arbitrary) initial phase of the
model is an additional parameter of the fit.

\begin{figure}
\begin{minipage}{95mm}
\includegraphics[width=90mm,angle=0]{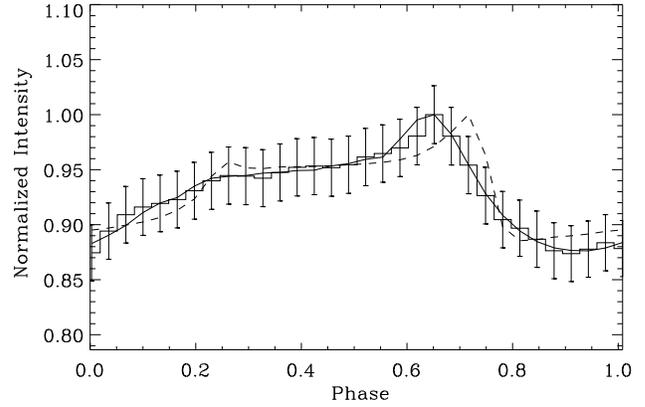}
\end{minipage}
\hfil\hspace{\fill}
\begin{minipage}{95mm}
\includegraphics[width=90mm,angle=0]{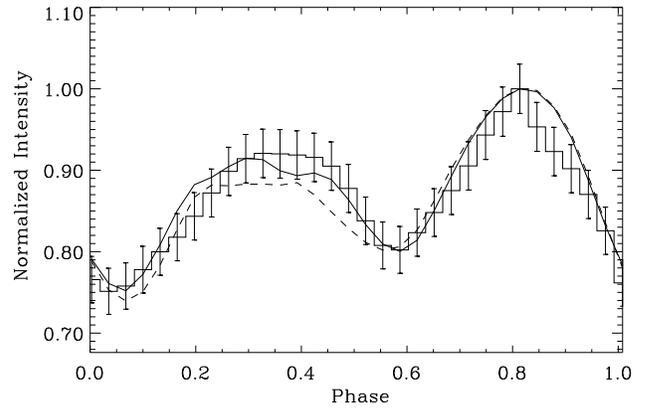}
\end{minipage}
\caption{\label{fit0806_1223} Same as in fig.~\ref{clustfit0420} for two
further XDINSs. Top:
the EPIC-PN
(0.12-1.2 keV) lightcurve
of RX J0806.4-4123 detected during rev.~618
(\citealt{hanoi2004}).
The best
fit parameters are: $b_0 = 0.39$,  $b_1 = -0.37$,  $b_2
= 0.12$,  $b_3 =
-0.13$,  $b_4 = 0.49$, $\xi = 0.0^\circ$,
$\chi=59.2^\circ$. Bottom: the EPIC-PN
(0.12-0.5 keV) lightcurve
of RBS 1223 detected during rev.~561
(\citealt{hab2003}). Best fit parameters are:
$b_0 = 0.21$,  $b_1 = 0.02$,  $b_2
= 0.59$,  $b_3 =0.53$,  $b_4 = 0.50$, $\xi = 0.0^\circ$,
$\chi= 95.1^\circ$}
\end{figure}

\begin{figure}
\begin{minipage}{95mm}
\includegraphics[width=90mm,angle=0]{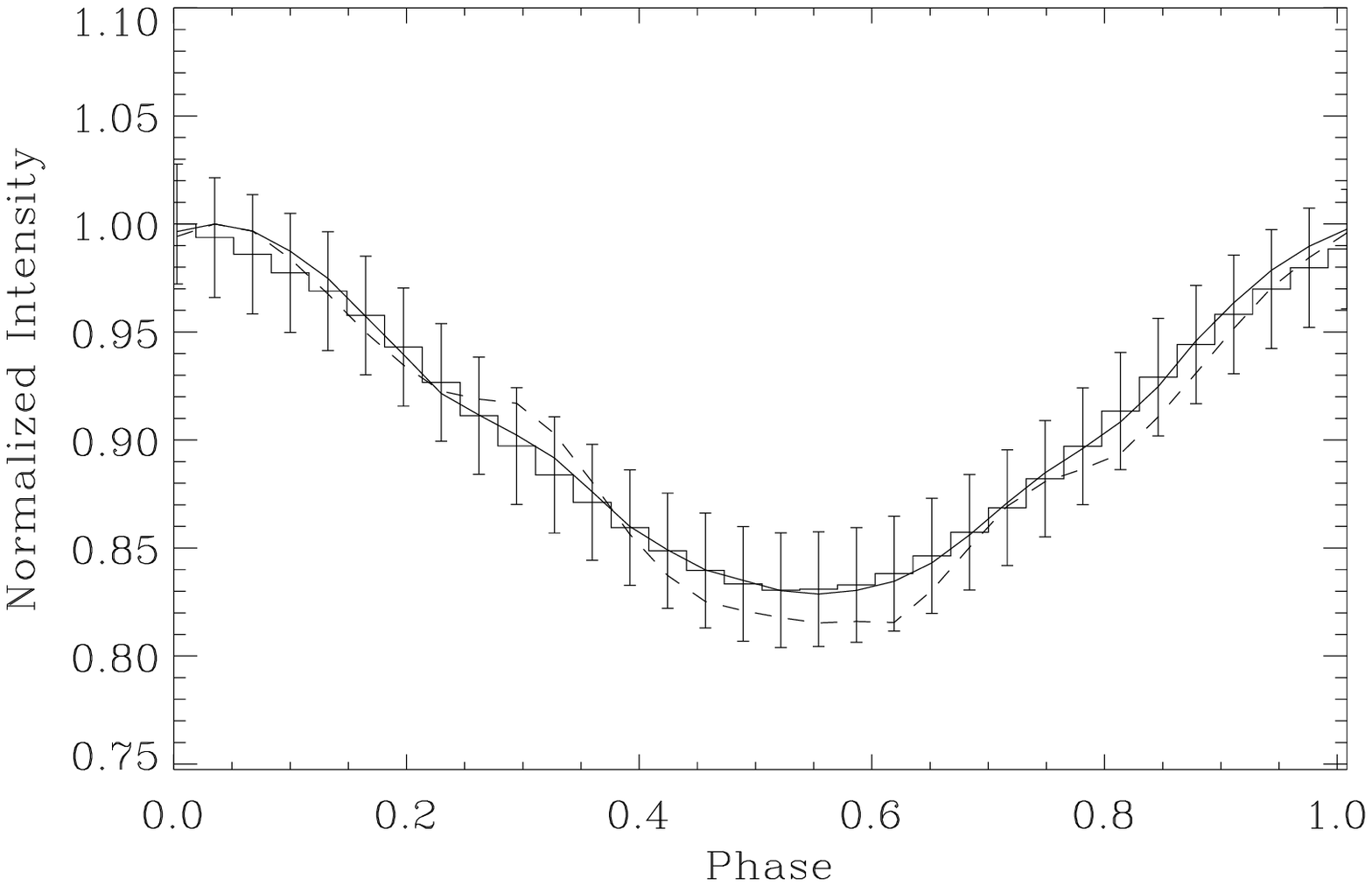}
\end{minipage}
\hfil\hspace{\fill}
\begin{minipage}{95mm}
\includegraphics[width=90mm,angle=0]{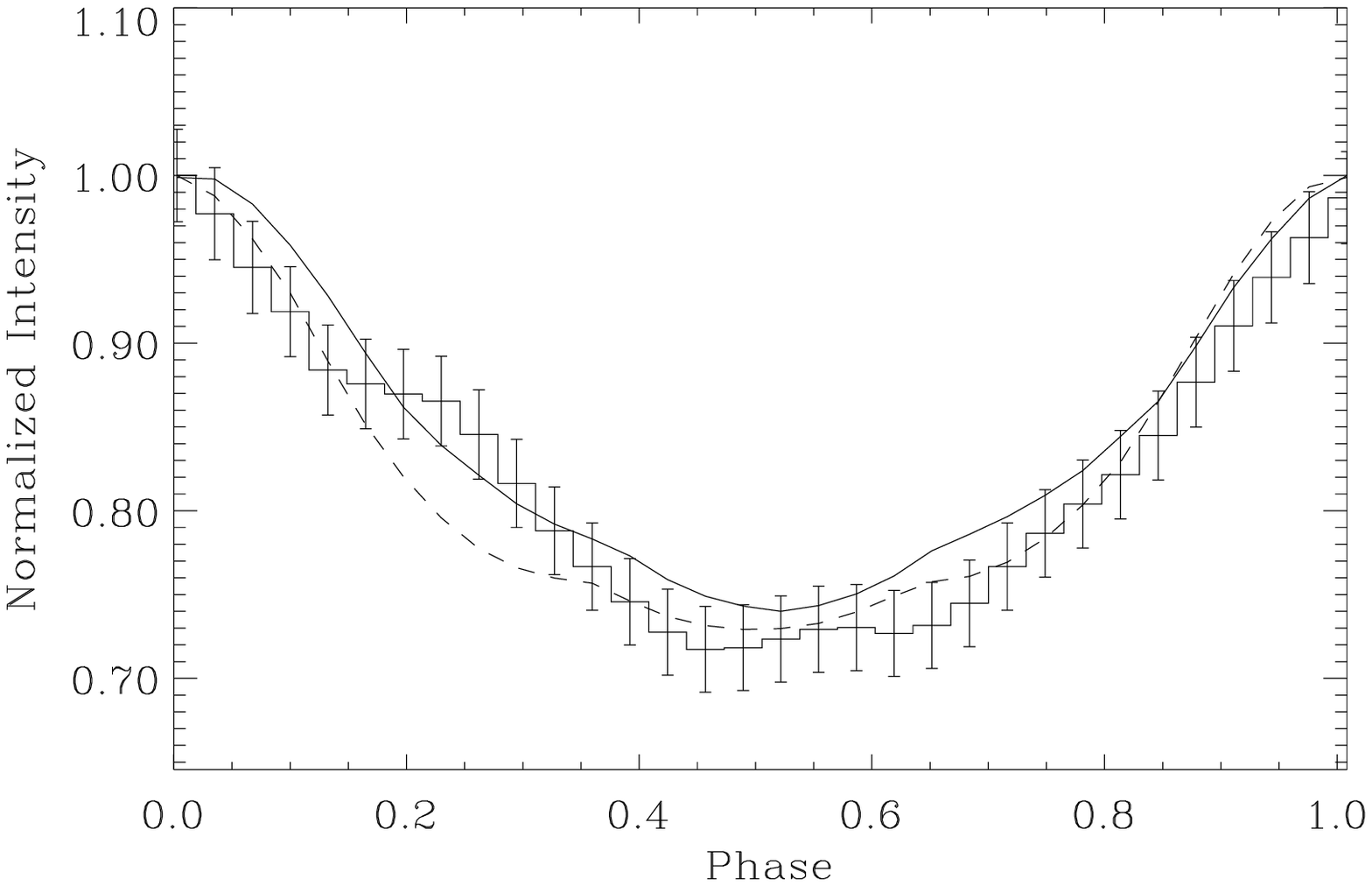}
\end{minipage}
\caption{\label{fit0720} Same as in fig.~\ref{clustfit0420} for two
different EPIC-PN
(0.12-1.2 keV) lightcurves detected from
RXJ0720.4-3125
(both from \citealt{devries04}). Top (rev.~78):
$b_0 = 0.36$,  $b_1 = 0.43$,  $b_2
= -0.16$,  $b_3 =
-0.16$,  $b_4 = -0.39$, $\xi = 0.0^\circ$,
$\chi= 68.1^\circ$. Bottom (rev.~711):
$b_0 = 0.45$,  $b_1 = 0.49$,  $b_2
= -0.06$,  $b_3 =
-0.08$,  $b_4 = -0.26$, $\xi = 0.0^\circ$,
$\chi= 87.6^\circ$. The second fit is obtained by leaving all
parameters free with respect to the solution  shown in the top panel.}
\end{figure}

The results of the fits are shown in figures \ref{clustfit0420},
\ref{fit0806_1223} and \ref{fit0720} for $B_{dip} = 6\times
10^{12}$~G, $\log T_p({\rm K}) = 6.1-6.2$. It is apparent that the
broad characteristics of all the XDINSs lightcurves observed so
far may be successfully reproduced for a suitable combination of
quadrupolar magnetic field components and viewing angles. However,
although in all cases a fit exists, we find that in general it is
not necessary unique. This means that the model has no
``predictive'' power in providing the exact values of the magnetic
field components and viewing angles. For this reason we do not
attempt a more complete fit, i.e. leaving also $T_p$ and $B_{dip}$
free to vary, nor we derive parameter uncertainties or confidence
levels. Our goal at this stage has been to show that there exist
at least one (and more probably more) combination(s) of the
parameters that can explain the observed pulse shapes, while this
is not possible assuming a pure dipole configuration.

\begin{figure}
\begin{minipage}{95mm}
\includegraphics[width=90mm,angle=0]{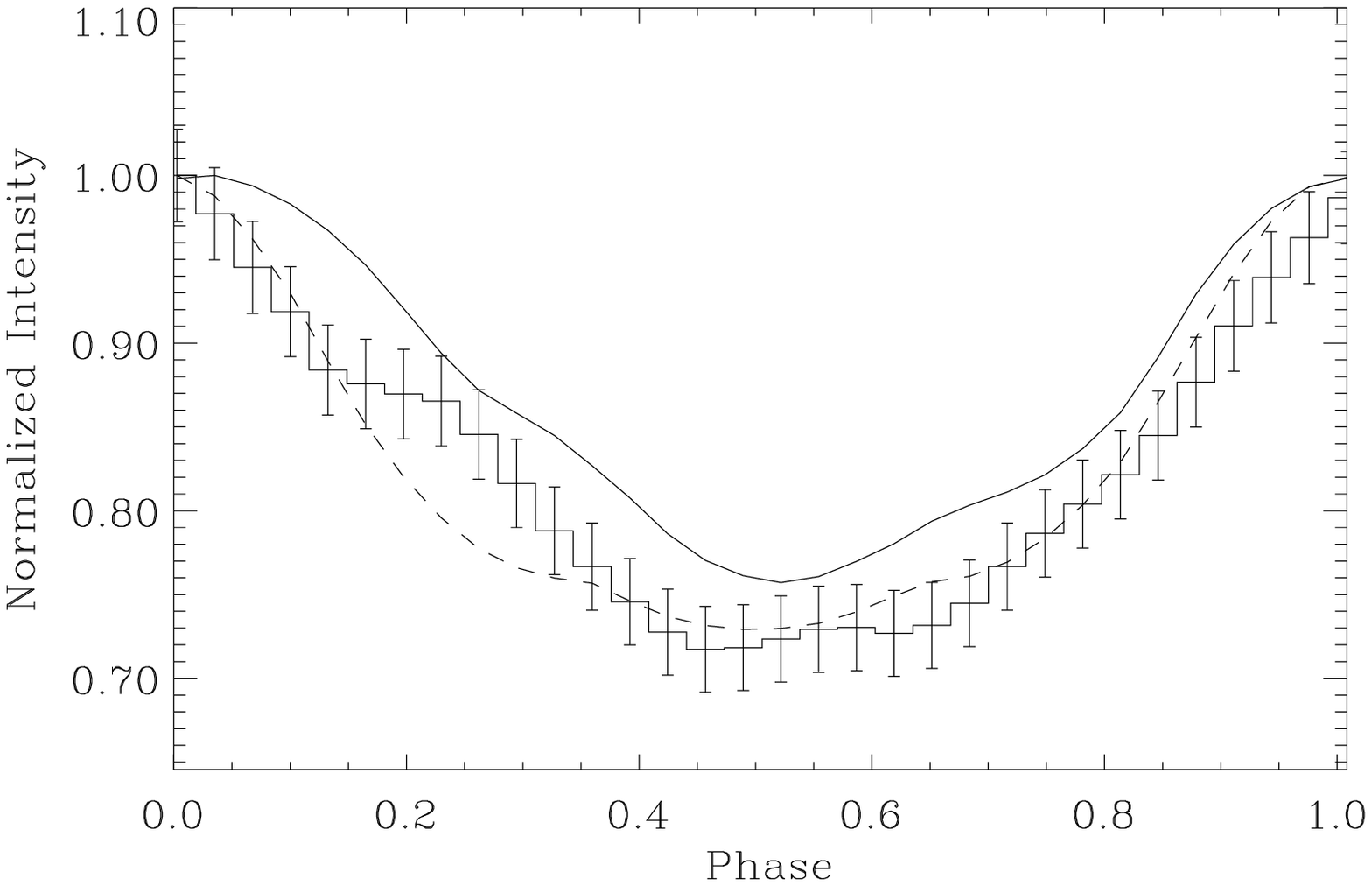}
\end{minipage}
\hfil\hspace{\fill}
\begin{minipage}{95mm}
\includegraphics[width=90mm,angle=0]{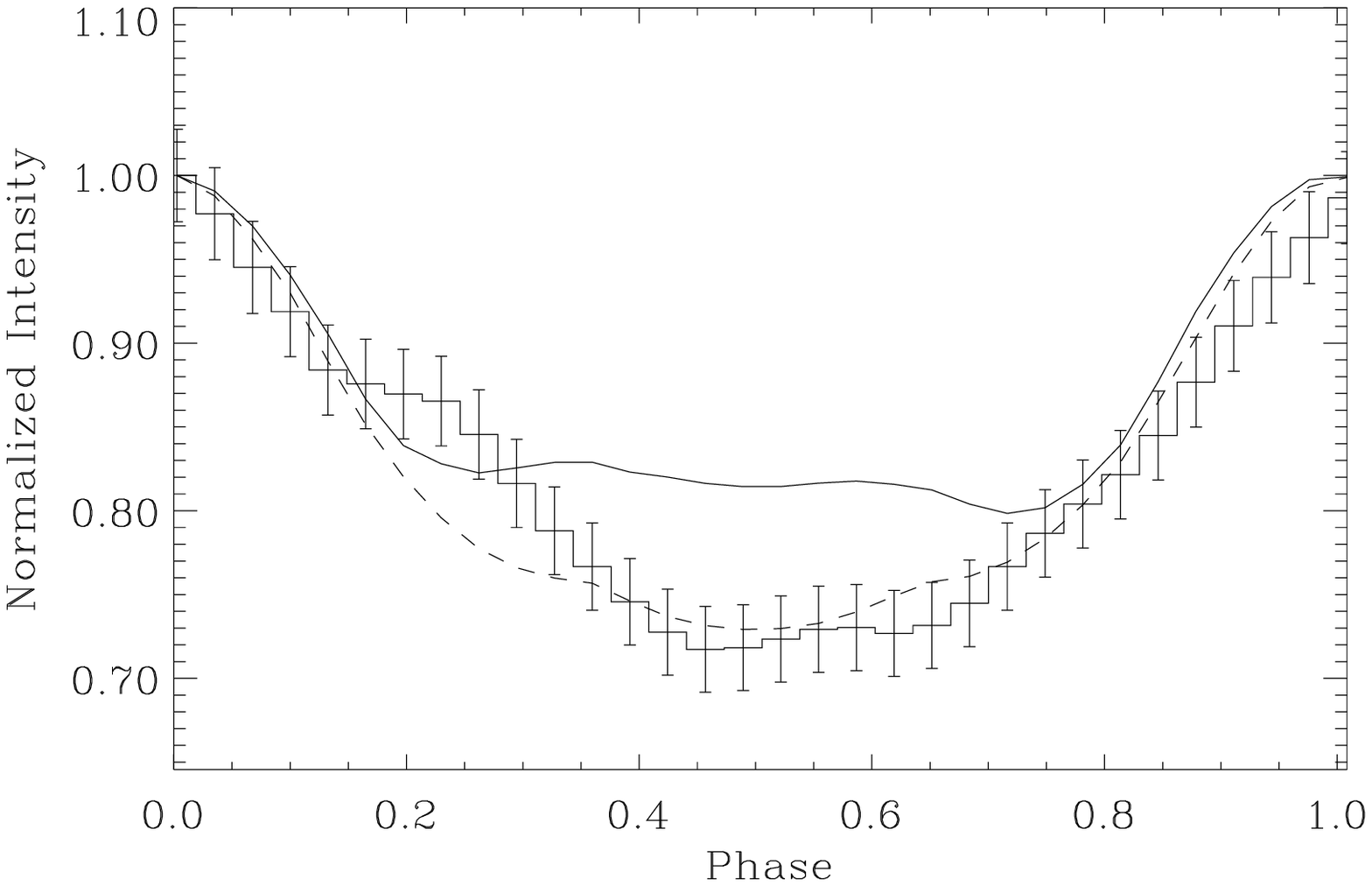}
\end{minipage}
\caption{\label{fit0720_noway} An attempt to fit the EPIC-PN
(0.12-1.2 keV) lightcurves detected from
RXJ0720.4-3125 during rev.~711 by using a fitting function with 
either the two viewing angles (top) or the values of $b_i$'s
($i=0...4$, bottom) kept fixed and equal to the values derived in the case 
of rev.~78. The dashed line is the same as in 
Fig.~\ref{fit0720}, lower panel. In this particular example, we also 
allowed a fractional change of $\sim20\%$ in the temperature between the 
two revolutions. Both attempts give an unsatisfactory result.}
\end{figure}

\begin{figure}
\begin{minipage}{95mm}
\includegraphics[width=90mm,angle=0]{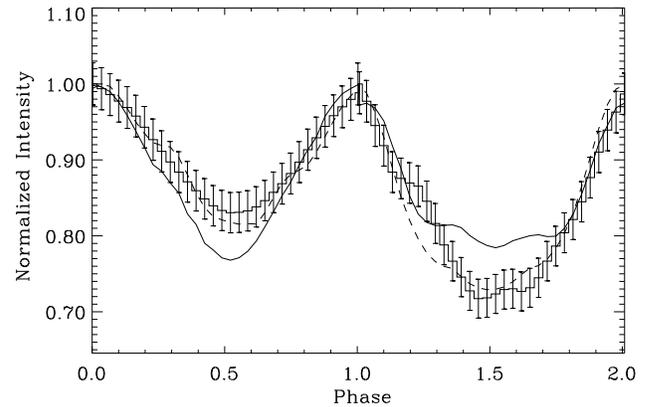}
\end{minipage}
\caption{\label{fit0720_sim} A simultaneous fit to both lightcurves
detected from RXJ0720.4-3125 during rev.~78 (shown in the phase 
interval 0-1) and rev.~711 
(phase 1-2). Only changes 
in the viewing angles have been allowed between the two epochs. The result 
is unsatisfactory.} \end{figure}

The case of RX J0720.4-3125 deserves, however, some further
discussion. This source, which was believed to be stationary and
as such included among {\it XMM-Newton\/} EPIC and RGS calibration
sources, was
recently found to exhibit rather sensible variations both in its
spectral and timing properties (\citealt{devries04};
\citealt{vink04}). In particular, the pulse shape changed over the
$\sim 2$~yrs time between {\it XMM\/} revolutions 78 and 711.
\cite{devries04} proposed that the evolution of RX J0720.4-3125
may be produced by a (freely) precessing neutron star. This
scenario can be tested by our analysis, since in a precessing NS
only the two angles $\xi$ and $\chi$ are expected to vary while
the magnetic field remains fixed. This means that having found one
combination of parameters which fits the lightcurve of rev. 78, a
satisfactory fit for rev. 711 should be obtained for the same
$b_i$'s and different values of the angles. Despite several
attempts, in which the proper values of $T_p$ as derived from the
spectral analysis of the two observations were used, we were
unable to obtain a good fit varying the angles only (see figure
\ref{fit0720_noway}). We performed
also a simultaneous fit to both lightcurves starting from a general
trial solution and asking that the $b_i$'s
are the same (but not necessarily coincident with those that best fit
data from rev.~78),  while the two angles (and the initial phases) are kept
distinct (see figure \ref{fit0720_sim}). Both approaches clearly indicate
that a change in both sets of quantities (magnetic field and angles) is
required (as in Fig.~\ref{fit0720}). From a physical point of view it is
not clear how magnetic field variations on
such a short timescale may be produced, therefore at present no
definite conclusion can be drawn. For instance, one possibility that
makes conceivable a change of
the magnetic field structure and strength on a timescale of years is that
the surface field is small scaled
(\citealt{gep03}). In this case, even small changes in the
inclination between line of sight and local magnetic field axis
may cause significant differences in the ``observed'' field
strength.

\section{Discussion}
\label{discuss}

X-ray dim isolated neutron stars (XDINs) may indeed represent
the Rosetta stone for understanding many physical properties of
neutron stars at large, including their equation of state.
Despite their potential importance, only recently detailed
observations of these sources have become progressively available
with the advent of  {\em Chandra} and {\em XMM-Newton} satellites.
These new data, while confirming the thermal, blackbody-like emission
from the cooling star surface, have revealed a number of spectral
and timing features which opened a new window on the study of these objects.
Some issues are of particular importance in this respect: i) the discovery
of broad absorption features at few hundreds eVs, ii) the quite
large pulsed fractions, iii) the departure of the pulse shape from a sine wave, and
iv) the long-term evolution of both the spectral and timing properties
seen in \zeroseven \ and, to some extent, in \zerofour. Pulse-phase
spectroscopy confirms that spectral and
timing properties are interwoven in a way which appears more
complex than that expected from a simple misaligned rotating dipole, as
the anti-correlation of the absorption line strength and of the hardness ratio
with the intensity along the pulse testify.

Motivated by this, we have undertaken a project aimed at studying
systematically the coupled effects of: i) radiative transfer in a
magnetized atmospheric layer, ii) thermal surface gradients,  and
iii) different topologies of the surface magnetic field in shaping the
spectrum and pulse shape. The ultimate goal of our investigation is
to obtain a simultaneous fit of both pulse profile and (phase-averaged
and -resolved) spectral distribution. As detailed comparisons
of synthetic spectra with observations have shown, no completely
satisfactory treatment of spectral modelling for these sources is available
as yet. For this reason, here we presented the general method
and concentrated on the study of the lightcurves, computed assuming a pure H,
magnetized atmosphere . The pulse shapes, in fact, should be less
sensitive on the details of the chosen atmospheric model.

We caveat the reader that our results have been computed
under a number of simplifying assumptions. For instance, there are still
considerable uncertainties in current modelling of
the NS surface thermal emission: we just mention here that most of
the observed NS spectra cannot be reproduced by the theoretical
models currently available (see \citealt{hab03}, \citealt{hab2004} for
reviews and references therein). Second, the surface
temperature has been
derived using eq. (\ref{tsurfgen}) which is based on the assumption
that the temperature profile is only dictated by the heat
transferred from the star interior. While this is expected to be
the main mechanism, other effects may significantly contribute.
For instance, heating of the star surface may occur
because of magnetic field decay, and the polar caps may be re-heated
by back-flowing particles or by internal friction. Third, we have computed
the emissivity by assuming that each atmospheric, plane parallel, patch
emits a total flux $\sigma T_s^4$. In other words, while the
spectral distribution
is computed using a proper radiative transfer calculation, we introduced
the further assumption that each slab emits the same total flux
as a blackbody radiator. A more consistent approach would be to
compute the spectrum emitted by each patch by taking the value of
$T_s$ and the efficiency of the crust as a radiator (\citealt{rob04})
as the boundary
condition at the basis of the atmosphere. Our
working assumption avoids the burden of creating a complete grid
of models in this parameter, with the disadvantage that the
spectral properties of each patch may be slightly approximated.

As far as the application to XDINSs is concerned, the greatest
uncertainties arise because in this paper we are not fitting
simultaneously the observed spectra and pulse shape. The
quadrupolar components and viewing angles are treated as free
parameters while the polar values $T_p$ and $B_{dip}$ are fixed
and, of course, they must fall in the domain spanned by our
atmosphere model archive. As stated earlier, albeit we have
checked that a fit is possible for different combinations of
$B_{dip}$ and $T_p$ in the allowed range, in general the archive
will not contain the exact values of $T$ inferred from spectral
observations of the coldest XDINSs. As long as we make the same
assumption that the local spectrum emitted by each patch of the
star surface is computed using fully ionized, magnetized hydrogen
atmosphere models, we do still expect to reach a good fit by using a
temperature of a few tens of eV smaller than
those used, albeit with different values of the quadrupolar components.
However, partial ionization effects are not included
in our work, and bound atoms and molecules can affect the results,
changing the radiation properties at relatively low $T$ and high
$B$ (\citealt{pocha03}; \citealt{pot2004}).

Even more crucial is the issue of fitting with a realistic value
of the magnetic field. The recent detection of absorption features
at $\approx 300$--700 eV in the spectrum of XDINSs and their
interpretation in terms of proton cyclotron resonances or
bound-bound transitions in H or H-like He, may indicate that these
sources possess strong magnetic fields, up to $B\sim 9 \times
10^{13}$~G (\citealt{vk2004}; \citealt{hanoi2004};
\citealt{sil05}). A (slightly) more direct measurement, based on the
spin-down measure, has been obtained so far
only in one source (i.e. \zeroseven, e.g. \citealt{cro2004},
\citealt{kmk04}). In this
case the measurement points at a more moderate field, $B\sim (2-3)\times
10^{13}$~G, which is only a few times larger than that used
here. However, the
possibility that (some) of these objects are ultra-magnetized NSs
is real. Would this be confirmed, our model archive should be
extended to include higher field values. This, however, poses a
serious difficulty, since the numerical convergence of model
atmospheres is particularly problematic at such large values of
$B$ for $T\la 10^6$~K. Moreover, as mentioned in
\S\ref{model_quad}, if such high field strengths are associated to
crustal (and not to star-centered) fields, the surface temperature
gradient is expected to be substantially different from that used
in the present investigation (\citealt{gkp04}).

The application presented here makes only use of the properties of
the pulse profile in the total energy band, and does not exploit
color and spectral information available for these sources. This
worsens the problem of finding a unique representation, problem
which is to some extent intrinsic, due to the multidimensional
dependence of the fitting function on  the various physical
parameters. Aim of our future work is to reduce the degeneracy by
refining the best fit solutions by using information from the
light curves observed in different color bands and/or from the
spectral line variations with spin pulse. Also, a more detailed
statistical analysis on the models population based on algorithms
more sophisticated than the PCA's \citep{gifi90,saeg05} may  shed light on
the possible  correlation between the physical parameters in
presence of strong non-linearity and possibly on the meaning of
the two subclasses identified through a cluster analysis.


\section{Acknowledgements}

We acknowledge D.A. Lloyd and R. Perna for allowing us to use
their radiative transfer code (which accounts for a non vanishing
inclination between the local
magnetic field and the local normal to the NS surface),
for their assistance in the set up
and for several discussions during the early stages of this
investigation.
Work partially supported by the Italian Ministry for Education,
University and Research (MIUR) under grant PRIN 2004-023189. SZ thanks
PPARC for its support through a PPARC Advanced Fellowship.


\clearpage

\begin{table*}
\begin{center}
\begin{tabular}{lcccl}
Isolated Neutron Stars Parameters.&&& \\
\hline \\
Source &
$P$ (s) & Semi-Ampl. &
XMM-Newton rev. &
Refs. \\
\hline \\
RX J0420.0-5022 & $3.45$ & $13\%$ & 570 & 1   \\
RX J0806.4-4123 & $11.37$ & $6 \%$ & 618 & 1   \\
RBS 1223 & $10.31$ &   $18 \%$ & 561 & 2  \\
\zeroseven   & $8.39$&  $11 \%$ & 78 & 3  \\
\zeroseven   & $8.39$ &  $16\%$   &  711 & 3  \\
RBS 1774  & $9.44$ &  $4 \%$ & 820 & 4 \\
\hline
\end{tabular}
\caption{The light curves used in
this paper are taken from references listed in the fourth column, which
correspond to:
[1] \citet{hanoi2004}; [2] \citet{hab2003}; [3] \citet{devries04}; [4]
\citet{sil05}.
}\label{tableins}
\end{center}
\end{table*}

\clearpage

\clearpage

\clearpage

\clearpage

\clearpage

\clearpage

\appendix
\section{}
\label{app1}

Let us write the magnetic field at any given point on the star surface in terms of its
cartesian components relative to the rotating frame $(x,\, y,\, z)$
\begin{equation}
\label{brot}
{\mathbf B}=B_x{\mathbf q}_\perp+B_y({\mathbf b}_{dip}\times{\mathbf q}_\perp)+
B_z{\mathbf b}_{dip}\, .
\end{equation}
The three unit vectors which identify the cartesian axes in the rotating frame have polar
components $r$, $\theta$, $\phi$ (relative to the same frame)
\begin{eqnarray}\label{rotpol}\nonumber
&&{\mathbf q}_\perp = (\sin\theta\cos\theta,\cos\theta\cos\phi,-\sin\theta\sin\phi)\\
\noalign{\smallskip}
&&{\mathbf b}_{dip}\times{\mathbf q}_\perp = (\sin\theta\sin\theta,\cos\theta\sin\phi,
\sin\theta\cos\phi)\\
\noalign{\smallskip}
&&{\mathbf b}_{dip} =(\cos\theta,-\sin\theta,0)\nonumber\, .
\end{eqnarray}
Taking the scalar product of ${\mathbf B}=(B_r,B_\theta,B_\phi)$ with each of the three
unit vectors above, we have
\begin{eqnarray}\label{bcartrot}\nonumber
&&B_x = B_r\sin\theta\cos\theta+B_\theta\cos\theta\cos\phi-B_\phi\sin\theta\sin\phi\\
\noalign{\smallskip}
&&B_y = B_r\sin\theta\sin\theta+B_\theta\cos\theta\sin\phi+B_\phi\sin\theta\cos\phi\\
\noalign{\smallskip}
&&B_z =B_r\cos\theta-B_\theta\sin\theta\nonumber\, .
\end{eqnarray}
The cartesian components of ${\mathbf B}$ in the fixed frame are then given by
\begin{eqnarray}\label{bcartfix}\nonumber
&&B_X = B_x({\mathbf q}_\perp)_X+B_y({\mathbf b}_{dip}\times{\mathbf q}_\perp)_X+
B_z({\mathbf b}_{dip})_X\\
\noalign{\smallskip}
&&B_Y = B_x({\mathbf q}_\perp)_Y+B_y({\mathbf b}_{dip}\times{\mathbf q}_\perp)_Y+
B_z({\mathbf b}_{dip})_Y\\
\noalign{\smallskip}
&&B_Z = B_x({\mathbf q}_\perp)_Z+B_y({\mathbf b}_{dip}\times{\mathbf q}_\perp)_Z+
B_z({\mathbf b}_{dip})_Z\nonumber\, .
\end{eqnarray}
\end{document}